\documentclass[aps,prd,showpacs,amsmath,amssymb,nofootinbib,eqsecnum, preprintnumbers,twocolumn]{revtex4-1}
 
\usepackage{hyperref} 
\usepackage{graphicx,xcolor}
\usepackage{epstopdf}

\newcommand{\be}{\begin{equation}}
\newcommand{\ee}{\end{equation}}
\newcommand{\ba}{\begin{eqnarray}}
\newcommand{\ea}{\end{eqnarray}}
\newcommand{\nn}{\nonumber} 
\newcommand{\cH}{\mathcal{H}}

\newcommand{\bb}[1]{\bar{#1}}

\begin{document}
\title{Conformal Triality of the Kepler problem}

\author{Marco Cariglia}
\email{marco@iceb.ufop.br}
\affiliation{Universidade Federal de Ouro Preto, ICEB, Departamento de F\'isica.
  Campus Morro do Cruzeiro, Morro do Cruzeiro, 35400-000 - Ouro Preto, MG - Brasil}

\date{\today}  

\begin{abstract} 
We show that the Kepler problem is projectively equivalent to null geodesic motion on the conformal compactification of Minkowski-4 space. This space realises the conformal triality of Minkwoski, dS and AdS spaces. 
\end{abstract}

\maketitle

\section{Introduction} 
The Kepler problem can be rightly considered as a 'golden classic' in Hamiltonian dynamics for a host of reasons. It is sufficiently realistic to describe within appropriate limits the dynamics of planets in the Solar System, and to some extent classical scattering of pointlike electric charges, while at the same time it is important in the theory of integrable systems, being  superintegrable and with a large dynamical symmetry group. It can be taught at a simple level in an undergraduate physics course of mechanics, as well as in a Quantum Mechanics course since integrability holds in the quantum theory too. At the same time a very extensive bibliography and books have been written on the subject and its generalisations, see  \cite{Cordani2003kepler} and related works. The literature on the subject being fairly vast, we do not try in this brief letter to cover it in a complete way but rather we try to point out the papers that are most relevant to the line of thought pursued here, and refer the reader to their bibliography to supplement what is missing. 
 
An important moment in the history of the problem comes with the modern, geometrical approach that can be attributed to Fock and then Bargmann \cite{Fock1935,Bargmann1936}, who analysed the quantum mechanical Kepler problem, and showed that the $SO(4)$ symmetry of negative energy states is explained by mapping the Kepler problem with a fixed energy to that of a free particle on a $3-$sphere embedded in $4-$dimensional space. The classical description has been discussed and then generalised by Moser and Belbruno. The former showed that the hypersurface of a given negative energy in phase space is homeomorphic to the unit tangent bundle of the sphere $S^3$ with the north pole excluded, the result being generalisable to general dimension $n$\cite{moser1970regularization}. Belbruno added the cases of positive energy, in correspondence to the three-hyperboloid $\mathcal{H}^3$, and zero energy which corresponds to 3-dimensional Euclidean space \cite{belbruno1977two}.

It is well known that, at the level of the allowed shapes of trajectories of the Kepler problem, ellipses, parabolae and hyperbolae are all related by projective transformations. Thus the question arises if it is possible to find a projective relation between different energy trajectories in the whole phase space. Recent work in the direction of relating geometrically trajectories of different energies 
has been done by Barrett, Keane and Simmons \cite{KBS2000}, who showed that the Kepler system is related, for all values of the energy at the same time, to geodesic motion on 3-spaces of constant curvature $k$ via canonical transformations and reparameterisation of trajectories. The relation introduced between the Kepler problem and the geodesic motion on constant curvature manifolds reminds somehow of coupling constant metamorphosis \cite{Hietarinta1984coupling}. The two systems described are considered by the authors to be related but different, their Hamiltonians not being the same. Depending on whether $k$ is positive, zero or negative one obtains geometrically the different dynamical symmetry algebras $so(4)$, $iso(3)$ and $so(3,1)$ of the Kepler problem. In \cite{KB2000} 4-dimensional Einstein static spacetimes are studied, which are foliated by the constant curvature 3-spaces mentioned above. These  spaces are conformally flat and their conformal group is $SO(4,2)$. Coordinates are found that put the metrics in a manifestly conformally flat form and the Lie algebra of conformal Killing vectors is studied. Null geodesic motion in these space is special since admits the maximum number of conserved charges that are linear in the momenta: they are generated by all the conformal Killing vectors. Keane \cite{Keane2002} generalised the constructions and described in detail the zero energy case and time dependent conserved quantities. 
  
The current work is motived from the desire to give a geometrical interpretation of the full  dynamical group $SO(4,2)$ of symmetries of the Kepler problem. An important role in discovering the nature of the above interpretation is played by projective transformations in phase space: as will be seen in the following section, these allow a simple and elegant reformulation of the classical Kepler problem in terms of null geodesics on a conformally flat Weyl space where the dynamical symmetry for all energies is manifest, thus displaying the strength of the technique used and completing the geometrisation of the system. 
Different conformal classes are projectively related, and in fact one is dealing with null geodesics on the space that realises the conformal triality of Mink, dS and AdS spaces \cite{Penrose1974,Cordani2003kepler,zhou2005conformal}. In many respects we follow a parallel line to that of Kean, Barrett and Simmons, the main difference being that we use projective geometry to show the dual systems of Kepler and geodesic motion are projectively the same, and  that the end result is null dynamics in a $4-$dimensional Weyl space that includes at the same time all the values of the curvature: this is the conformal compactification of Minkowski space and as shown it is related to the null lift of Kepler's problem in phase space. In particular, trajectories with different energy sign can be mapped one into the other via a projective transformation. Previous treatments only considered the action of $SO(2,4)$ on standard Kepler trajectories, which does not change the sign of the energy. We also show how to embed trajectories of the original Kepler problem in the space $\mathbb{R}^{2,4}$ which is used to describe the conformal compactification of Minkowski space, realising geometrically the action of $SO(2,4)$ on the projective trajectories.

\section{Main result\label{sec:main}} 
The Hamiltonian of the Kepler problem is 
\be \label{eq:H} 
H = \frac{1}{2} p_i p_i - \frac{\alpha}{q} \, , 
\ee 
where $\{q^i, p_j \}$ are conjugate variables, $i,j=1,2,3$, $q = \sqrt{q^i q^i}$ and for simplicity in this work we exclude the configuration space point $q = 0$. $\alpha > 0$ is a constant. We consider the following null lift of the Hamiltonian: 
\be \label{eq:null_first}
\mathcal{H} = \frac{1}{2} p_i p_i - \frac{p_y^2}{q} - p_a^2 + p_b^2 \, , 
\ee 
where we have added new conjugate variables $\{y, p_y \}, \{a, p_a \}$, $\{b, p_b\}$, the momenta being conserved. This is a null Hamiltonian that projects to the original one if we impose $\mathcal{H} = 0$, $p_y^2 = \alpha$, and then $p_a^2 = E$, $p_b = 0$ for positive energy solutions of \eqref{eq:H}, $p_a = 0$, $p_b^2 = - E$ for negative energy solutions, and $p_a = 0$, $p_b = 0$ for zero energy solutions. For reasons that will become clear in a few moments, we choose to work in the open phase space region $p_y > 0$. We can exclude $p_y = 0$ since its associated geodesics are related to free motion as it can be seen from \eqref{eq:null_first}. 
 
In \cite{KBS2000} the following dual Hamiltonian is considered 
\be \label{eq:G} 
G = \frac{1}{4} \left(k + \frac{p_i p_i}{2} \right)^2 q^2 \, , 
\ee 
which can be canonically transformed into a geodesic Hamiltonian. As the authors note, $G$ above can be formally obtained by solving for $\alpha$ in \eqref{eq:H}, taking its square, setting $-E = k$, and promoting the result to a phase space function. This is quite similar to what happens in coupling constant metamorphosis \cite{Hietarinta1984coupling}, with the exception of the procedure of taking the square which is novel. 
 
We now relate \eqref{eq:H} and \eqref{eq:G} projectively in phase space. As we have shown in a recent companion work \cite{MarcoProjective2015} any null geodesic Hamiltonian of the form $\frac{1}{2} g^{AB}(q) p_A p_B$, as is $\mathcal{H}$ above, defines a projective conic in tangent space, and any two lower dimensional systems whose null lifts are related one to the other by a projective transformation are dual. In particular, one can rescale the metric $g_{AB}$ using a phase space factor $\Omega^2(q,p) \neq 0$ and obtain a new Hamiltonian 
\be \label{eq:tilde_H} 
\bb{\cH} = \frac{1}{2} \Omega^{-2} (q,p) g^{AB} \, p_A p_B \, .  
\ee 
If $\bb{\cH}$ generates evolution of trajectories associated to a parameter $\bb{\lambda}$, and $\cH$ to $\lambda$, then then the equations of motion for the variables $(q,p)$ given by $\bb{\cH}$ are mapped back to the original equations of motion of $\cH$ after one makes the following change of parameter on one trajectory at a time 
\be 
d\bb{\lambda} = \Omega^{2}(q(\lambda),p(\lambda)) \, d\lambda \, , 
\ee
the result being valid for null geodesics only. We refer the reader to \cite{MarcoProjective2015} for details. 
 
In the present situation we will perform two such reparameterisations. First choose the phase space factor to be $\Omega^{-2} = q$: we change parameter $\lambda \rightarrow \lambda^\prime$ with 
\be 
d\lambda^\prime = \frac{1}{q(\lambda)}  \, d\lambda \, , 
\ee 
which transforms ${\cH}$ into 
\be 
\mathcal{H}^\prime = \left( \frac{1}{2} p_i p_i  - p_a^2 + p_b^2 \right) q   - p_y^2 \, . 
\ee 
This reparameterisation is valid only for trajectories that have $q \neq 0$, so we are implicitly excluding collision trajectories. The same change of parameter appears   in \cite{KBS2000}, as well as other previous works in the literature, but not in the sense of acting on a null lift. 
 
 
We now perform a non-trivial, phase space dependent rescaling of $\mathcal{H}^\prime$ times the factor $\Omega^{-2} =  \left( \frac{1}{2} p_i p_i  - p_a^2 + p_b^2 \right) q   + p_y^2$: this gives a new Hamiltonian 
\be 
\bb{\cH} = \left( \frac{1}{2} p_i p_i  -p_a^2 + p_b^2 \right)^2 q^2   - p_y^4 \, , 
\ee 
where 
\be \label{eq:Hbar}
d\bb{\lambda} = \left[  \left( \frac{1}{2} p_i p_i  - p_a^2 + p_b^2 \right) q(\lambda^\prime)   + p_y^2 \right]^{-1} \, d\lambda^\prime \, .  
\ee 
The reader can notice that this is a null lift of \eqref{eq:G} upon identifying $k = - E$. 
The rescaling factor is never zero since from the condition $\mathcal{H}^\prime = 0 = \mathcal{H}$ we get $p_y^2 = \alpha = \left( \frac{1}{2} p_i p_i  - p_a^2 + p_b^2 \right) q(\lambda^\prime)$. Now we perform the following canonical transformation: 
\ba 
Y &=& \frac{y}{\sqrt{2} p_y} \, , \nn \\ 
p_y^2 &=&  \sqrt{2} P_y  \, , 
\ea 
which is defined in the open region $P_y > 0$. This has the effect of replacing the $p_y^4$ term in \eqref{eq:Hbar} with $2 P_y^2$. Next, we perform the canonical transformation of \cite{KBS2000}: 
\ba 
q^i &=& \frac{1}{2\sqrt{2}}\left( Q^2 P_i - 2 (Q \cdot P ) Q^i \right) \, , \\ 
p_i &=& - 2 \sqrt{2} \frac{Q^i}{Q^2} \, , 
\ea 
complemented by $A = p_a$, $P_A = - a$, $B = p_b$, $P_B = - B$. 
This transforms the Hamiltonian $\bb{\mathcal{H}}$ into 
\be \label{eq:null_H_final}
\bb{\mathcal{H}} = 4 \left[ \frac{1}{2} \left(1 - \frac{(A^2 - B^2) Q^2}{4} \right)^2 P_i P_i - \frac{P_y^2}{2} \right] \, , 
\ee 
where $A^2 - B^2 =  E$. 
This null Hamiltonian can be interpreted in two ways. On one hand it is the null lift associated to positive energy geodesic motion on a 3-dimensional Riemannian manifold that is conformally flat, of constant curvature $k = - E$, written in stereographic coordinates, as noticed in \cite{KBS2000}. There the authors state that the geodesic motion Hamiltonian and the Kepler problem Hamiltonian \eqref{eq:H} are related but different Hamiltonians, however in our null projective approach we have just shown that they are obtained from the same projective null Hamiltonian and therefore represent the same dynamics. 
On the other hand, the null Hamiltonian \eqref{eq:null_H_final} describes null geodesics on an Einstein static spacetime that is conformally Minkowskian, as discussed in \cite{Keane2002,KB2000}. Thus the present approach encompasses both types of discussions. 
We also notice that in our framework what matters is not the specific metric but its Weyl class. The Weyl space we have found is the conformally flat 4-dimensional space that hosts the conformal triality of Minkowski, de Sitter and anti-de Sitter spaces \cite{zhou2005conformal}, that will be discussed in more detail in the next section. 
 
We now briefly discuss conserved quantities. Suppose that $C(q,p)$ is a conserved quantity for a null geodesic Hamiltonian $\mathcal{H}$, $\{C , \mathcal{H} \} = 0$ where $\{ \cdot, \cdot \}$ is the Poisson bracket. It is clear that, in our projective approach, $C$ will also be conserved for a rescaled Hamiltonian $\bb{\mathcal{H}} = \Omega^{-2}(q,p) \mathcal{H}$, using the on-shell condition $\mathcal{H} = 0$. From this we can infer that conserved quantities for the null Hamiltonians \eqref{eq:null_first}, \eqref{eq:null_H_final} can be obtained from those of flat Minkowski space. In particular, the conformal group $SO(2,4)$ of flat Minkowski space acts as a symmetry algebra for $\bar{\mathcal{H}}$. 
The explicit change of variables that conformally maps the Einstein static metric of \eqref{eq:null_H_final} into Minkowski space can be found in \cite{KB2000}. In the next section we will see how the action of $SO(2,4)$ on the projective curves described in this section can be described geometrically by mapping these curves into curves in a projective cone in a six-dimensional ambient space.

\section{Conformal triality} 
We are dealing with a Weyl 4-dimensional space that includes Mink, dS and AdS spaces. The fact that these are conformally related, together with the fact that the Lie algebra of the Kepler problem is $so(2,4)$ have a profound common origin. It is known in fact that the conformal group of 4-dimensional Minkowski space is locally isomorphic to $O(2,4)$, and to our knowledge the first discussion of the origin of these results is due to \cite{Penrose1974}. Similar arguments have been discussed among others by \cite{Cordani2003kepler,zhou2005conformal}, and in particular we will refer to the former for a detailed discussion. We remind the reader of  the original arguments and then show how they apply to the present problem. 
One can consider $\mathbb{R}^{2,4}$ with coordinates $(T,V, W, X_1, X_2, X_3)$ and Lorentzian metric $\eta = \text{diag}(-1,-1,1,1,1,1)$. The null cone $N$ given by points that satisfy  
\be \label{eq:null_cone} 
- T^2 - V^2 + W^2 + X_1^2 + X_2^2 + X_3^2 = 0 
\ee 
is left invariant by the action of $O(2,4)$. If we consider the intersection of $N$ with the null hyperplane $V-W=1$ then we can use $(T,X_1, X_2, X_3)$ as coordinates, and the induced metric on the intersection is 
\be 
ds^2 = - dT^2 + dX_1^2 + dX_2^2 + dX_3^2 \, . 
\ee 
Thus we can identify the intersection with 4-dimensional Minkwowski space, $Mink_4$. Now consider the set $[N]$ of unoriented lines on $N$, i.e. sets of points identified under the relation $(T,V, W, X_1, X_2, X_3) \sim \lambda (T,V, W, X_1, X_2, X_3)$ for $\lambda \neq 0$, and such that \eqref{eq:null_cone}  holds. Then, every element of $[N]$ that can represented by coordinates with $V \neq W$ will intersect the hyperplane $V-W=1$ in a unique point, and can be put in a one to one correspondence with a point in $Mink_4$. However, there are also elements of $[N]$ with coordinates such that $V=W$, and these never intersect $Mink_4$. These elements, according to \eqref{eq:null_cone}, describe a light cone in $Mink_4$ and therefore can be considered points at infinity. The result is that $[N]$ can be thought of as $Mink_4$ with its points at infinity added. We notice that points on $N$ can be described by the coordinates 
\ba \label{eq:N_coordinates}
&& T = r \cos y \, , \quad V = r \sin y \, , \nn \\ 
&& (W,X_1, X_2, X_3) = r V^\alpha (\theta^i) \, , 
\ea 
where $\alpha = 1, \dots, 4$, $i=1,2,3$, and $V^\alpha$ satisfies $V^\alpha V^\alpha = 1$, and thus is a vector on $S^3$ parameterised by coordinates $\theta^i$. From this we see that $(y, \theta^i)$ can be taken as local coordinates for $[N]$, which is then homeomorphic to $(S^1 \times S^3)/ \mathbb{Z}_2$. Being a compact space, it is known as the \textit{conformal compactification} of $Mink_4$. 
 
While there is no natural metric on $[N]$, we can consider a set of representative points on $N$ by taking a (local) cross-section of $N$ given by points of the type \eqref{eq:N_coordinates} with $r = r (y, \theta^i)$. Then we can consider the induced metric on the cross section, which is given by 
\be 
ds^2 = r^2 (y, \theta^i) \left( - dy^2 + ds^2_{S^3} \right) \, , 
\ee 
i.e. it is conformal to a Robertson-Walker spacetime with constant positive curvature on the equal time slices, using the same variables $(y, \theta^i)$ defined above. Since such space is conformally flat, this includes the Minkowski space described earlier. Then $[N]$ is a conformal manifold, that is a manifold equipped with a metric defined modulo conformal rescalings: different cross sections correspond to different rescalings.  
 
Going back to the results of the previous section, for each choice of the evolution parameter $\lambda$ \eqref{eq:null_H_final} describes null geodesic motion on one possible cross-section, and vice-versa: the independence on the choice of parameter points out to the existence of a curve on $[N]$ that is independent on the cross-section used. We now want to define a more precise correspondence between the null geodesics of \eqref{eq:null_H_final} and curves on $[N]$, and we will do this by first embedding the former into special curves on $N$ as follows.  Consider a curve on $N$ with equation 
\be \label{eq:curve}
\lambda \mapsto r(\lambda) P(\lambda) \, , 
\ee 
where $P(\lambda)$ is the embedding of $S^1\times S^3 \subset \mathbb{R}^{2,4}$ given by \eqref{eq:N_coordinates} for $r=1$. The vectors $\frac{\partial P}{\partial y}$, $\frac{\partial P}{\partial \theta^i}$ form a basis for the tangent space of $S^1\times S^3$, and together with $P$ they form a basis for the tangent space of $N$. We will define geodesics on $[N]$, in the sense of zero acceleration curves, by asking that the second derivative of \eqref{eq:curve} has zero components along $\frac{\partial P}{\partial y}$ and $\frac{\partial P}{\partial \theta^i}$: in other words, the acceleration can only have a component along the vector $P$ itself. Then one finds the following equation for $x^\mu = (y, \theta^i)$ 
\be 
\ddot{x}^\mu + \Gamma^\mu_{\rho\sigma} \dot{x}^\rho \dot{x}^\sigma +  \frac{d \ln r^2}{d\lambda} \dot{x}^\mu \, , 
\ee 
where $\Gamma^\mu_{\rho\sigma}$ are the Christoffel symbols of the induced Robertson-Walker metric for $r=1$, and $r$ is an arbitrary function of $\lambda$. This is a geodesic equation on $S^1\times S^3$ with a non-affine parameter, where the affine parameter $\mu$ is recovered using $\frac{d}{d\mu} = r^2 \frac{d}{d\lambda}$. Since these are unparameterised null geodesics of $S^1\times S^3$, then they can be mapped into the geodesics of \eqref{eq:null_H_final}. Summarising, we have proven  that null geodesics of \eqref{eq:null_H_final}, which are the Kepler motions identified under projective transformations, are in one-to-one correspondence with geodesics on $[N]$. 
 
We make a final observation. It seems from \eqref{eq:null_H_final} that the null lift manifold of the original Kepler problem corresponds locally to the the six dimensional space described in this section. For one might use the coordinate $A^2 - B^2$ to parameterise which type of section of $N$ one is taking, whether it is 
the intersection with a sphere,  a hyperboloid, or a hyperplane, together with the radius of 
curvature of the former two, and the coordinates $y$, $Q_i$ to describe the section. Lastly one could use a different combination of $A$, $B$ to parameterise the direction perpendicular to 
the null cone $N$. This seems locally correct but whether, or how, it can extended to a global result is a different statement that we do not investigate here. We find it however suggestive. 
 
Concluding, we believe that the technique used here, projective transformations in phase space, with the vast amount of freedom associated to it, will stimulate further research in the field of Hamiltonian dynamics and integrable systems.


\section*{Acknowledgments}
\noindent The author would like to thank C. M. Warnick for reading the manuscript and useful comments. Funding by PROPP, Federal University of Ouro Preto is acknowledged. 


\bibliographystyle{apsrmp}
\bibliography{Bibliography_06Apr2015}

\end{document}